**Ultrahigh Green and Red Optical Gain Cross-sections from Solutions of Colloidal Quantum Well Heterostructures**

Savas Delikanli,[†] Onur Erdem,[†] Furkan Isik, Hameed Dehghanpour Baruj, Farzan Shabani, Huseyin Bilge Yagcı and Hilmi Volkan Demir*

Dr. S. D. and Prof. H. V. D.
Luminous! Center of Excellence for Semiconductor Lighting and Displays, School of Electrical and Electronic Engineering, Nanyang Technological University, 50 Nanyang Avenue, Singapore 639798, Singapore

Dr. S. D. and Prof. H. V. D.
Division of Physics and Applied Physics, School of Physical and Mathematical Sciences, Nanyang Technological University, 21 Nanyang Link, Singapore 639798, Singapore

Dr. S. D., Dr. O. E., F. I., H. D. B., F. S., H. B. Y. and Prof. H. V. D.
 Department of Electrical and Electronics Engineering, Department of Physics, UNAM – Institute of Materials Science and Nanotechnology, Bilkent University, Ankara 06800, Turkey

*Corresponding author: hvdemir@ntu.edu.sg; volkan@bilkent.edu.tr

[†] These authors contributed equally to this work

heterostructures, colloidal quantum wells

**Abstract**

Optical gain in solution, which provides high photostability as a result of continuous regeneration of the gain medium, is extremely attractive for optoelectronic applications. Here, we propose and demonstrate amplified spontaneous emission (ASE) in solution with ultralow thresholds of 30 $\mu J/cm^2$ in red and of 44 $\mu J/cm^2$ in green from engineered colloidal quantum well (CQW) heterostructures. For this purpose, CdSe/CdS core/crown CQWs, designed to hit the green region, and CdSe/CdS@$Cd_xZn_{1-x}S$ core/crown@gradient-alloyed shell CQWs, further tuned to reach the red region by shell alloying, were employed to achieve high-performance ASE in the visible. The net modal gain of these CQWs reaches 530 $cm^{-1}$ for the green and 201 $cm^{-1}$ for the red, two orders of magnitude larger than those of colloidal quantum dots (QDs) in solution owing to intrinsically larger gain cross-sections of these CQWs. To explain the root cause for ultrahigh gain coefficient in solution, we show that for the first time that the gain cross-sections of these CQWs is $\geq 3.3 \times 10^{-14}$ $cm^2$ in the green and $\geq 1.3 \times 10^{-14}$ $cm^2$ in the red which are two orders magnitude larger compared to those of CQDs. These findings confirm the extraordinary prospects of these solution-processed CQWs as solution-based optical gain media in lasing applications.

## Introduction

Colloidal quantum wells (CQWs) present an excellent platform as gain media owing to their solution processability, wide tunability of emission colour,[1-4] giant oscillator strength,[1,5] large gain cross-section,[6] slow Auger rates[7-9] and ultra-narrow emission profiles.[1,3,10] Especially the giant gain cross-section and slow Auger recombination of CQWs make them extremely attractive candidates for optical gain studies and applications as these are the critical factors for satisfying the gain condition.[11] Lasing and amplified spontaneous emission (ASE) with ultralow thresholds[2,7,8,12-15] and large net modal gains[7,14,16] from solid films of close-packed type-I and type-II CQWs in different heterostructures, including the architectures of core/crown, core/shell and core/crown/shell, have been widely studied. Although solution-based gain medium can be highly advantageous because of the enhanced photostability owing to constant replenishment of gain media by the flux of nanocrystals and flexibility of incorporation into optical cavities, there are only very few reports of optical gain in solution using colloidal nanocrystals.[6,17-20] This is primarily due to the low concentration of gain media possible in solution, limiting the feasible levels of gain coefficients. For this reason, almost all of the previous optical gain demonstrations using CQWs and other colloidal nanocrystals were achieved by using of their solid films with high packing-density. While thin solid films of nanocrystals may lead to lower thresholds and larger modal gains due to the increased density of gain media, such films are more likely to suffer from surface roughness affecting the waveguiding and optical confinement,[21,22] the optical scattering as a result of microscopic aggregation and stacking of particles,[23] nonradiative losses due to Förster resonance energy transfer (FRET) and homo-FRET,[24,25] and reproducibility of the identical samples. In addition, the usage of solid films of colloidal nanocrystals under gain usually requires extra processes, such as encapsulation with a glass slide or a barrier polymer, to prevent them from degradation in the ambient atmosphere under constant excitation.[12,26]

In-solution optical gain can be achieved straightforwardly by employing a simple glass tube[17] or a cuvette[6,18] serving possibly as a cavity and a host along with highly concentrated nanocrystal solution serving as a gain medium. In this approach, uniform gain media in a cavity can be achieved robustly and reproducibly, and loss mechanisms, such as nonradiative homo-FRET and scattering of light due to aggregations of nanocrystals, are conveniently suppressed. Such a gain medium can be utilized in microfluidic networks to achieve on-chip lasing for optical sensing and detection.[27-31] Despite the foreseeable advantages of CQWs, including the modal gain reaching levels one order of magnitude larger than those of the quantum dots and nanorods,[16] slow Auger rates[7-9,32] and gain cross-sections with three orders of magnitude larger than the quantum dots,[6] there is only one previously reported work on CQW based optical gain in solution to date which was achieved by multiphoton pumping using a cuvette-based Fabry–Perot resonator.[6] The achieved lasing by two- and three-photon pumping was shown to be highly stable and polarized, confirming the prospects of heterostructures of CQWs in solution-based optical gain.[6]

Here, different than the previous works, we present the achievement of in-solution ASE from green-emitting CdSe/CdS core/crown CQWs and red CdSe/CdS@Cd$_x$Zn$_{1-x}$S core/crown@gradient-alloyed shell CQWs enabling ultralow thresholds of 44 µJ/cm$^2$ and 30 µJ/cm$^2$, respectively, in solution. In addition, we measured net modal gains of these CQWs having a concentration of ~1.7×10$^{16}$ cm$^{-3}$ corresponding to a molarity level of ~28.3 µM in solution using variable stripe length (VSL) method. The ASE and VSL measurements were performed using a glass capillary tube. The net modal gains of green CdSe/CdS core/crown CQWs and red CdSe/CdS@Cd$_x$Zn$_{1-x}$S core/crown@gradient-alloyed shell CQWs at this concentration reach ~530 cm$^{-1}$ and ~201 cm$^{-1}$, respectively. While these net modal gains are two to three orders of magnitude larger than those of QDs in solution,[17] they still closely match the net modal gains of core/shell CQWs in solution,[6] showing the prospects of these CQWs as a solution-based gain media. In addition, we report for the first time the gain cross-section of these heterostructures of CQWs , which are found to be $\geq 3.3 \times 10^{-14}$ cm$^2$ for the green-emitting CdSe/CdS core/crown CQWs and $\geq 1.3 \times 10^{-14}$ cm$^2$ for the red-emitting CdSe/CdS@Cd$_x$Zn$_{1-x}$S core/crown@gradient-alloyed shell CQWs. The obtained gain cross-sections are two orders of magnitude larger than those of QDs[11,17] which can be attributed to the high density of states in CQWs. These findings show the vast potential of CQWs as in-solution optical gain media.

**Methods**

For the purpose of systematic study of optical gain in solution, we synthesized CdSe/CdS core/crown CQWs[33] to hit the green spectral range and CdSe/CdS@Cd$_x$Zn$_{1-x}$S core/crown@gradient-alloyed shell CQWs[34] to reach the red range, both using high quality 4 monolayer (ML) CdSe CQWs as the seeds. In both cases CdS crown was grown on these starting seed CdSe CQWs using a previously reported recipe.[33] Here, the recombination takes place in the CdSe core of these core/crown CQWs due to the large valence band offset between CdSe and CdS layers and large exciton binding energy. Transmission electron microscopy (TEM) images of the CdSe core CQWs and CdSe/CdS core/crown CQWs are shown in Figures 1(a) and 1(b), respectively. The lateral area was measured to be ~150 nm$^2$ for the core CQWs and ~350 nm$^2$ for the core/crown CQWs.

To further obtain the CdSe/CdS@Cd$_x$Zn$_{1-x}$S core/crown@gradient-alloyed shell CQWs, 4 ML thick gradient alloyed shell of Cd$_x$Zn$_{1-x}$S was grown on the previously synthesized CdSe/CdS core/crown CQWs to push the emission peak towards red.[34] The lateral area of these core/crown@gradient-alloyed shell CQWs is 450 nm$^2$ while the thickness of these CQWs having 4 MLs of CdSe and 8 MLs of CdZnS layers is ~4.25 nm.[7] The CQWs were dispersed in toluene for optical characterization and optical gain measurements. Details of the synthesis of CQWs are provided in the Supporting Information (SI). Owing to the gradient alloyed shell of Cd$_x$Zn$_{1-x}$S, charge carriers feel soft potential confinement in the vertical dimension[34], which is also highly desirable for optical gain applications due to the suppression of Auger recombination.[35] In this heterostructure with gradient alloyed shell, electron wavefunction leaks largely into the shell region due to the shallow conduction band-offset, whereas the hole wavefunction is mostly localized in the core as a result of the large valence band-offset between CdSe core and $Cd_xZn_{1-x}S$ shell.[34] A representative TEM image of these core/crown@gradient-alloyed shell CQWs is presented in Figure 1(c).

The absorption profiles of the CdSe core, CdSe/CdS core/crown and CdSe/CdS@$Cd_xZn_{1-x}S$ core/crown@gradient-alloyed shell CQWs are given in Figure 1(d). The absorption cross-sections of CdSe/CdS core/crown and CdSe/CdS@$Cd_xZn_{1-x}S$ core/crown@gradient-alloyed shell CQWs are $1.1\times10^{-13}$ cm$^2$ and $2.9\times10^{-13}$ cm$^2$, respectively, at 400 nm, which were calculated using a method provided in our previous publication.[36] The intrinsic absorption coefficients $\mu_i(\lambda)$ of the CdSe/CdS core/crown and CdSe/CdS@$Cd_xZn_{1-x}S$ core/crown@gradient-alloyed shell CQWs, which are obtained by dividing the absorption cross-sections at the wavelength of interest by the physical volume of the CQWs, are $1.96\times10^5$ cm$^{-1}$ and $2.1\times10^4$ cm$^{-1}$, respectively, at their heavy hole absorption peak.[36] It is worth mentioning that the intrinsic absorption is an important parameter showing the gain capability of a material since the material gain $g(\lambda)$ at a specific wavelength is equal to the intrinsic absorption coefficient $\mu_i(\lambda)$ of the material at the same wavelength.[37] The photoluminescence (PL) spectra of CdSe/CdS core/crown and CdSe/CdS@$Cd_xZn_{1-x}S$ core/crown@gradient-alloyed shell CQWs in solution are shown in Figure 1(e). The PL peak of the core/crown CQWs is at ~517 nm (in solution) with a full-width-half-maximum (FWHM) of 8 nm while that of CdSe/CdS@$Cd_xZn_{1-x}S$ core/crown@gradient-alloyed shell CQWs (in solution) is at ~642 nm with a FWHM of ~21 nm. The narrow emission profile of CQWs is due to their magic size thickness, which inhibits inhomogeneous broadening stemming from size dispersion.

To explore the ASE performance of our CQWs in solution, the CQWs dispersed in toluene were inserted into a capillary tube having an inner diameter of 300 µm by the help of capillary force. The higher boiling point (compared to most of non-polar solvents, such as commonly used hexane), the lower volatility and the relatively high refractive index of toluene make it our choice of solvent. The low volatility of toluene suppresses the chance of particle aggregation as a result of the solvent evaporation under photoexcitation by the pump laser. In addition, the higher refractive index of toluene compared to most commonly used hexane helps confining the optical field within the gain medium more efficiently, thereby resulting in higher modal optical gain. The concentration of the CQWs in the capillary tube is calculated to be ~$1.7 \times 10^{16}$ cm$^{-3}$. The capillary tubes filled with CQW solution were pumped at 400 nm using femtosecond pulses with a 120 fs pulse width and at a 1 kHz repetition rate. Excitation was performed with a stripe geometry using a cylindrical lens. ASE signal was collected from the edge of the sample by coupling the emitted photons into an optical fiber connected to a spectrometer. Figure 2(a) shows pump-fluence dependence of the ASE spectra from CdSe/CdS core/crown CQWs. The sharp ASE peak can be observed at 534.2 nm with a FWHM of 4.3 nm at room temperature on the red side of the spontaneous emission for the pump fluences above the threshold. This red shift of ASE ($\approx 5$ nm) with respect to the spontaneous emission confirms multiexcitonic gain, which is usually seen in type-I nanocrystals.[38] This sort of red shift in ASE is desirable for optical gain since this hinders self-absorption.[7,39] Total emission intensity as a function of the pump fluence for the green-emitting CQWs is presented in Figure 2(b). Using this data, the ASE threshold is calculated to be 44 µJ/cm$^2$, which is lower than any previously reported ASE threshold in solution for nanocrystals. Previously, the best reported threshold in solution for colloidal nanocrystals is ~105 µJ/cm$^2$ from CsPbBr$_3$ perovskite nanocrystals.[18]

We used variable-stripe-length (VSL) method[40] to determine the modal gain of CQWs in solution at various pump fluences higher than the optical gain threshold. Figure 2(c) displays total emission intensities at different pump fluences as a function of the stripe length. For the numerical fitting, we employed a fitting function of

$$I = A \times (e^{g_{net}L} - 1)/g_{net} \tag{1}$$

where $g_{net}$ is the net modal gain coefficient, $A$ is a parameter associated with spontaneous emission intensity, and $L$ is the length of the excitation stripe.[40] Figure 2(d) presents the pump-dependent net modal gain coefficient at various pump fluences obtained from the fittings presented in Figure 2(c). As can be seen in Figure 2(d), the net modal gain increases linearly at lower fluences above the threshold and saturates around ~530 cm$^{-1}$. [7,16] This relatively high gain coefficient in solution[17] can be attributed to the higher number of density of states in CQWs[6] and high concentration of CQWs in toluene. Although these net modal gains are nearly one order of magnitude lower than their solid films[7,16] due to lower concentration of CQWs, these net modal gains are similar to the net modal gains of solid films of quantum dots and nanorods[41-44] and two to three orders of magnitude larger than that of QDs in solution.[17]

The gain cross-section per CQW ($\gamma_g$) can be calculated by

$$g_{net} = C \times \varGamma \times \gamma_g - \alpha \tag{2}$$

where $C$ is the concentration of CQWs, $\varGamma$ is the optical confinement factor, and $\alpha$ is the optical loss coefficient.[6] In our case, C is ~1.7×10$^{16}$ cm$^{-3}$ in toluene, $g_{net}$ is known at different pump fluences as presented in Figure 2(d), and $\varGamma$ is calculated by numerically solving the supported optical modes, as will be discussed next. Here, we assume that the net modal gain is close to the modal gain ($g$) for optical gain in solution.[6] In fact, our net modal gain levels (530 cm$^{-1}$ for CdSe/CdS core/crown CQWs and 201 cm$^{-1}$ for CdSe/CdS@Cd$_x$Zn$_{1-x}$S core/crown@ shell CQWs) are much larger than even the loss coefficient from films of CQWs (~10 cm$^{-1}$),[22] which are expected to suffer larger loss in film compared to in solution. In addition, we added term $\varGamma$ into the equation (2) to account for the optical confinement factor in the active optical gain media. When the area of the pump beam is not much larger than the probe, the optical confinement factor in the active optical gain maybe different than unity and need to be carefully checked for the calculation of $\gamma_g$.

Here, the optical confinement factor ($\Gamma$) was calculated by obtaining the electrical field distribution in our system. $\Gamma$ for a mode is defined as the ratio of the modal power inside the excited portion of the gain media to the total modal power:

$$\Gamma = \int_{\substack{\text{excited} \\ \text{gain region}}} \frac{1}{2} \text{Re}\{\vec{E} \times \vec{H}^*\}.\hat{z}\mathrm{dxdy} \; / \int_{-\infty}^{\infty} \frac{1}{2} \text{Re}\{\vec{E} \times \vec{H}^*\}.\hat{z}\mathrm{dxdy} \tag{3}$$

where $\vec{E}$ and $\vec{H}$ are the electric and magnetic fields of the specified mode, respectively. This calculation is based on the fact that amplified spontaneous emission in the active gain medium is only possible for the modes that have large spatial overlap with the excitation and that other modes fail to supply sufficient gain exceeding the losses. A minimum value for the gain cross-section $\gamma_g$ can be found by computing $\Gamma$ for the modes that are guided and confined. The distributions of guided power in the hybrid $HE_{vm}$ modes were obtained using MATLAB. In our calculations, we used a refractive index of 1.51 as the effective refractive index of CQW solution and a refractive index of 1.46 for SiO$_2$ layer, which is assumed to extend to infinity. This is justified for the bound modes in the active region since the surrounding SiO$_2$ layer is optically thick and does not alter the mode profiles considerably (see Supporting Information for further details). The modal power distributions for the first four modes ($HE_{11}$, $HE_{21}$, $HE_{31}$, $HE_{12}$) for CdSe/CdS core/crown CQWs are presented in Figure 3(a-d). TE, TM and EH modes are not shown, since far from the cut-off TE/TM modes are approximated by $HE_{2m}$ modes and $EH_{v,m}$ modes are approximated by $HE_{v-2,m}$. The confinement factor peaks at 0.94 for $HE_{11}$ due to strong overlap with the excitation. Based on this maximal $\Gamma$, we computed the minimum gain cross-section of CdSe/CdS core/crown CQWs using the modal gain value of 530 cm$^{-1}$ and the concentration of $1.7 \times 10^{16}$ cm$^{-3}$. The obtained value of gain cross-section for CdSe/CdS core/crown CQWs is $\geq 3.3 \times 10^{-14}$ cm$^2$, which is 2 orders of magnitude larger than the gain cross-section of QDs[17] having gain profiles saturating approximately twice of their thresholds, suggesting a limited number of carrier species contributing gain in the case of QDs. We attribute this extremely large gain cross-section of CQWs to the high number of density of states in the CQWs.

Figure 4(a) presents the pump-fluence dependence of ASE spectra from the solution of CdSe/CdS@Cd$_x$Zn$_{1-x}$S core/crown@gradient-alloyed shell CQWs in a capillary tube pumped at 400 nm using femtosecond pulses (120 fs). In these CQWs, the spontaneous emission peaks at 651.9 nm with a FWHM of 29.0 nm. As the pump intensity increases, a sharp peak emerges at 653.0 nm with a FWHM of 6.0 nm and ultimately surpasses the PL spectrum at stronger pump fluences. The reduced red shift of ASE in this sample compared to that of the core/crown one can be attributed to the multiexcitonic gain with quasi-type-II band alignment profile. Total emission intensity as a function of the pump fluence is given in Figure 4(b). The ASE threshold for these red-emitting CQWs is 30 µJ/cm$^2$, which is lower than that of the green core/crown CQWs. This reduction in the gain threshold can be attributed to the further efficient suppression of Auger recombination in these alloyed-shell CQW heterostructures compared to the core/crown CQWs as a result of the graded alloying providing a soft confinement potential.[7,35] Figure 4(c) shows total emission intensities at different pump fluences as a function of the stripe length and the net modal gain coefficients at various pump fluences obtained from the fittings given in Figure 4(c) are presented in Figure 4(d). As shown in Figure 4(d), the net modal gain saturates around 201 cm$^{-1}$ at a pump fluence of ~350 µJ/cm$^2$. This maximum net modal gain from these red CQWs in solution is two orders of magnitude higher than that of CdZnS/ZnS alloyed-core/shell QDs[17] and similar to that of core/shell CQWs.[6] Modal characteristics of the red-emitting CQWs are identical to CdSe/CdS core/crown CQWs due to the large fiber parameter $V$ and similar refractive index of CQW solutions, and hence both CQW solutions share the same $\Gamma$ value of 0.94 at the ASE peak wavelength. For these red CQWs, the gain cross-section is found to be $\geq 1.3 \times 10^{-14}$ cm$^2$, which is again two orders of magnitude larger than that of QDs.[11] However, the gain cross-section of these core/crown@gradient-alloyed shell CQWs is lower than that of core/crown CQWs ($\sim 3.3 \times 10^{-14}$ cm$^2$). This is also reflected by the lower absorption coefficient of the core/crown@gradient-alloyed shell CQWs compared to the core/crown CQWs. This can be explained by to the quasi type-II band alignment of thicker core/crown@gradient-alloyed shell CQWs, which effectively reduces the oscillator strength of the transitions in the core/crown@gradient-alloyed shell CQWs. On the other hand, the intriguingly lower threshold of the core/crown@gradient-alloyed shell CQWs is dominantly dictated by their larger absorption cross-section at the excitation wavelength and stronger suppression of Auger recombination as compared to the core/crown CQWs.

**Conclusion**

In summary, we demonstrated ASE in solution using CdSe/CdS core/crown CQWs designed to emit in green and CdSe/CdS@Cd$_x$Zn$_{1-x}$S core/crown@gradient-alloyed shell CQWs tuned to emit in red dispersed in toluene with ultralow thresholds of 44 µJ/cm$^2$ and 30 µJ/cm$^2$, respectively. The net modal gain coefficient of these green and red reaches $\sim 530$ cm$^{-1}$ and $\sim 201$ cm$^{-1}$, respectively, at a concentration of $\sim 1.7 \times 10^{16}$ cm$^{-3}$, which are two to three orders of magnitude larger than that of QDs in solution.[17] The lower ASE threshold of CdSe/CdS@Cd$_x$Zn$_{1-x}$S core/crown@gradient-alloyed shell CQWs compared to CdSe/CdS core/crown CQWs is due to the larger absorption cross-section of the core/crown@gradient-alloyed shell CQWs at the pump wavelength. In addition, we found the gain cross-section of core/crown and core/crown@gradient-alloyed shell CQWs for the first time. The gain cross-sections of the green CdSe/CdS core/crown CQWs and the red CdSe/CdS@Cd$_x$Zn$_{1-x}$S

core/crown@gradient-alloyed shell CQWs are $\geq 3.3 \times 10^{-14}$ cm$^2$ and $\geq 1.3 \times 10^{-14}$ cm$^2$, respectively. The obtained gain cross-sections are two orders of magnitude larger than the gain cross-section of QDs and can be attributed to the higher available number of density of states in CQWs. These solution-processed CQWs with their exceptional properties as a solution-based optical gain medium presents extraordinary opportunities for the design and implementation of solution-based lasers, which can be integrated conveniently and intimately into microfluidic devices designed for sensing and imaging applications.

**Acknowledgements**

S.D. and O.E. contributed equally to this work. The authors gratefully acknowledge the financial support in part from Singapore National Research Foundation under the programs of NRF-NRFI2016-08, NRF-CRP14-2014-03 and the Science and  the Singapore Agency for Science, Technology and Research (A*STAR) SERC Pharos Program under Grant No. 152-73-00025 and in part from TUBITAK 115F297, 117E713, and 119N343. H.V.D. also gratefully acknowledges support from TUBA. O.E. acknowledges support from TUBITAK BIDEB.

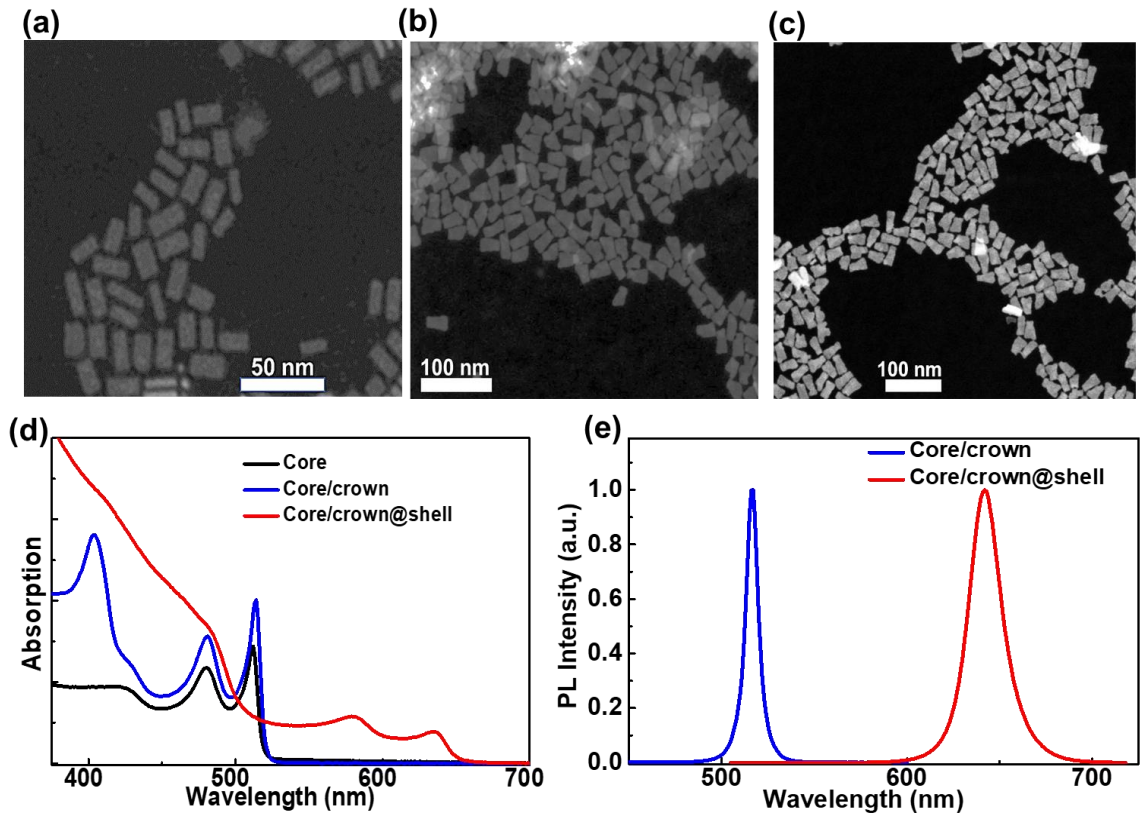

**Figure 1.** TEM images of (a) 4 ML CdSe, (b) 4 ML CdSe/CdS core/crown and (c) 12 ML CdSe/CdS@Cd$_x$Zn$_{1-x}$S core/crown@gradient-alloyed shell CQWs. (d) Absorption spectra of CdSe, CdSe/CdS core/crown and CdSe/CdS@Cd$_x$Zn$_{1-x}$S core/crown@gradient-alloyed shell CQWs. (c) PL spectra of CdSe/CdS core/crown and CdSe/CdS@Cd$_x$Zn$_{1-x}$S core/crown@gradient-alloyed shell CQWs.

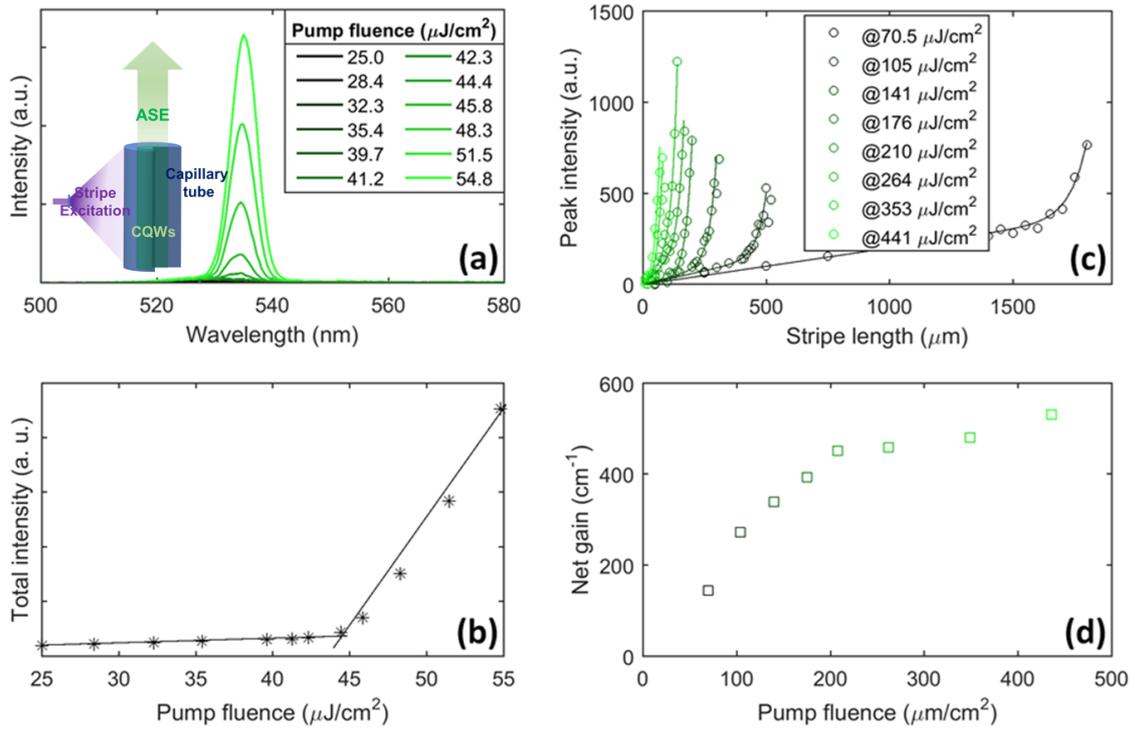

**Figure 2.** Optical gain performance of in-solution CdSe/CdS core/crown CQWs. (a) Collected optical spectra at different pump fluences. Inset of (a) shows a schematic of CQW solution in the capillary tube optically pumped using stripe geometry for ASE measurements. (b) Emission intensity as a function of the pump fluence. (c) Emission intensity as a function of the excitation stripe length. (d) Extracted net gain coefficients as a function of the pump fluence.

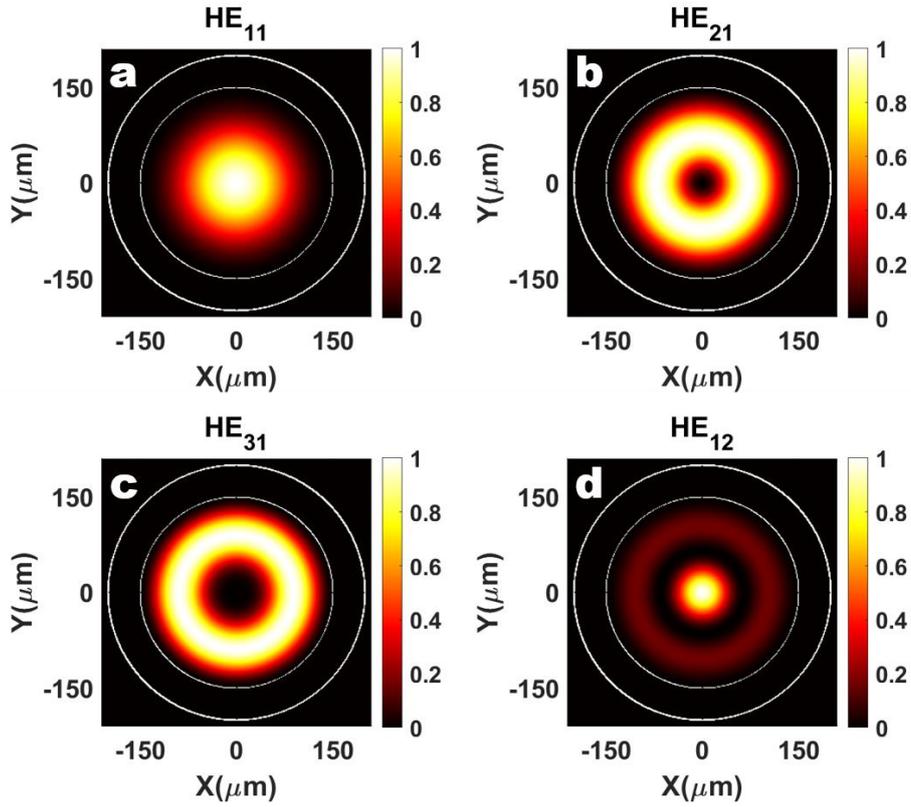

Figure 3. (a-d) Modal guided power distributions at 534 nm for the first four modes. The boundaries of the fiber are indicated with white concentric rings with large ring (r = 200 µm) corresponding to the air-cladding boundary and small ring (r = 150 µm) corresponding to cladding-core boundary. Due to the large fiber parameter $V$ and a similar refractive index of our CQW solutions, the calculated profiles at 653 nm and 534 nm are practically identical. The reason behind the near-unity confinement factor for $HE_{11}$ is easily understood from the regional overlap between mode and the excitation (see Figure S2).

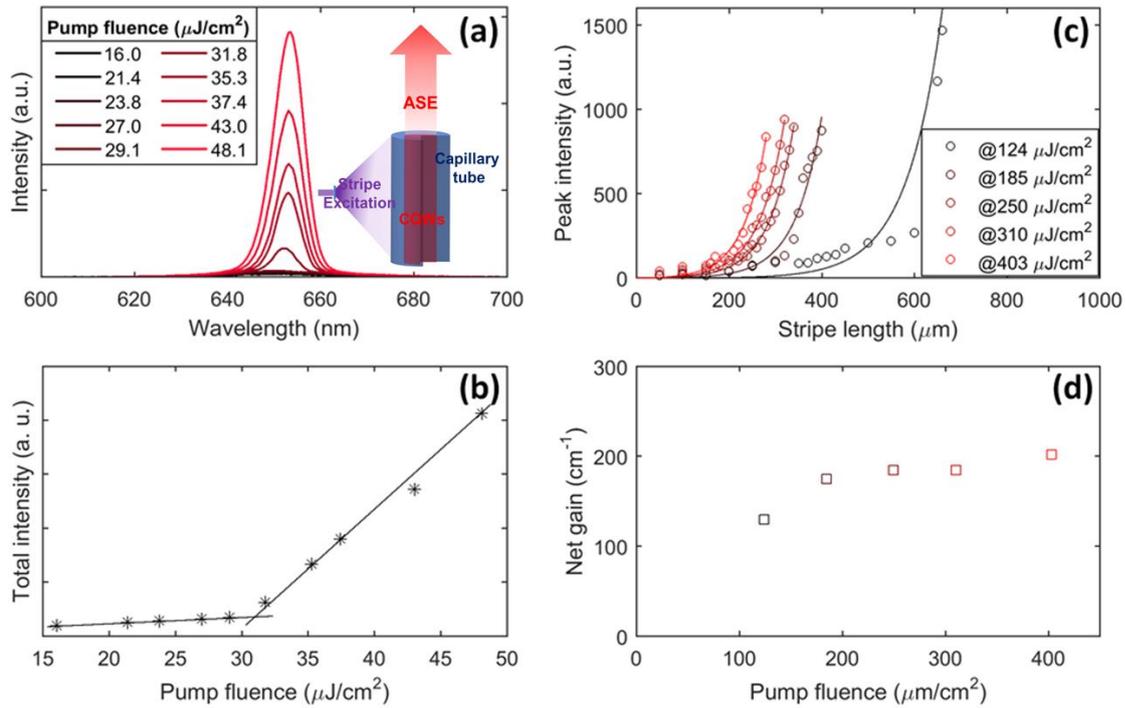

**Figure 4.** Optical gain performance of in-solution CdSe/CdS@Cd$_x$Zn$_{1-x}$S core/crown@gradient-alloyed shell CQWs. (a) Collected optical spectra at different pump fluences. Inset of (a) displays a schematic of CQW solution in the capillary tube pumped using stripe geometry for ASE measurements. (b) Emission intensity as a function of the pump fluence. (c) Emission intensity as a function of the excitation stripe length. (d) Extracted net gain coefficients as a function of the pump fluence.

**Green and red amplified spontaneous emission (ASE) in solution with ultralow thresholds of 44 μJ/cm² and of 310μJ/cm²**, respectively, from colloidal quantum well (CQW) heterostructures is demonstrated. The gain cross-sections of green and red CQWs is $\geq 3.3 \times 10^{-13}$ cm² and $\geq 1.1 \times 10^{-13}$ cm², respectively, two orders magnitude larger than those of quantum dots. These findings support the extraordinary prospects of these-solution processed QWs as solution-based optical gain media in lasing applications.

S. D., O. E., F. I., H. D. B. F. S., H. B. Y. and H. V. D.*

**Ultrahigh Green and Red Optical Gain Cross-sections from Solutions of Colloidal Quantum Well Heterostructures**

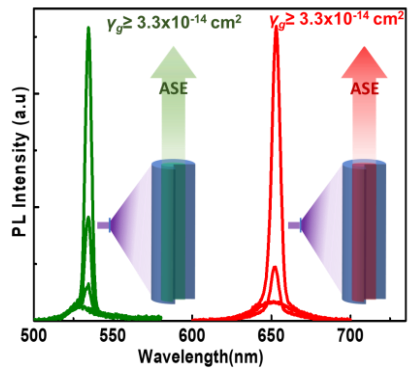

# Supporting Information

**Ultrahigh Green and Red Optical Gain Cross-sections from Solutions of Colloidal Quantum Well Heterostructures**

Savas Delikanli,[†] Onur Erdem,[†] Furkan Isik, Hameed Dehghanpour Baruj, Farzan Shabani, Huseyin Bilge Yagcı and Hilmi Volkan Demir*

**Chemicals:** Sodium myristate, cadmium acetate dihydrate (Cd(OAc)$_2 \cdot$ 2H$_2$O), cadmium nitrate tetrahydrate (Cd(NO$_3$)$_2$.4H$_2$O), zinc nitrate hexahydrate (Zn(NO$_3$)$_2$.6H$_2$O), oleylamine (OLA), N-Methylformamide (NMF), ammonium sulfide solution (40-48 wt. % in water), oleic acid (OA), selenium, hexane, acetonitrile, toluene, chloroform and ethylenediamine (EDA) were purchased from Sigma Aldrich .

**Synthesis of cadmium myristate:** Cadmium myristate was synthesized according to a previously published protocol with slight modifications.[1] In a typical synthesis, 6.26 g of sodium myristate and 2.46 g of cadmium nitrate tetrahydrate were separately dissolved in 500 and 80 mL of methanol, respectively. After complete dissolution, they were mixed together and stirred vigorously for 5 h. Finally, the white cadmium myristate powder was precipitated by centrifugation. In order to remove any undesired impurities and side products, the precipitate was washed couple of times with methanol. The final product was dried overnight under vacuum at room temperature (RT) and stored under ambient conditions.

**Synthesis of CdSe core of CQWs:** 4 monolayer (ML) CdSe CQWs were synthesized using a former procedure reported in the literature.[1] 68 mg of Se powder and 800 mg of cadmium myristate were added to 60 mL of 1-octadecene (ODE) in a 100 mL three-neck flask and degassed at 95 °C for 1 h. Then, the temperature was set to 240 °C under the argon flow. At 195 °C, 200 mg of Cd (OAc)$_2 \cdot$2H$_2$O was added to the reaction. This solution was kept at 240 °C for 8 min. Finally, the growth was terminated by adding 4 mL of oleic acid and quenched with a cold-water bath. Monodisperse 4 ML CdSe CQWs were obtained through size selective precipitation of the reaction mixture. The final product was stored in hexane.

**Synthesis of growth precursor:** A previously reported procedure was followed with slight modifications.[1] 480 mg of cadmium acetate dihydrate (Cd $(OAc)_2\cdot 2H_2O$), 340 µL of OA, and 2 mL of ODE were loaded in a beaker. The solution was sonicated for 30 min at room temperature under ambient atmosphere. By continuous stirring at 160 °C and sonication at 100 °C alternately, a mixture of white, homogeneous gel was obtained. After cooling the sample to the room temperature, 3 mL of 0.1 M sulfur in ODE was added to the mixture. The final solution was kept under constant stirring.

**Synthesis of CdSe/CdS core/crown CQWs:** A typical core-seeded synthesis method reported previously was used with slight modifications.[1] 4 ML CdSe core CQWs dispersed in hexane and 15 mL of ODE were introduced in a three-neck flask. The solution was degassed at 90 °C for 1 h. Then, the growth solution was injected to the mixture at 90 °C and degassed for 1 h. After that, the solution was heated to 240 °C under argon flow and held at this temperature for 5 min to ensure the proper growth of crown. The reaction was quenched with cold-water bath. The synthesized core/crown CQWs were precipitated by addition of 5 mL hexane and 3 mL ethanol and centrifugation. Finally, the precipitate was dispersed and stored in hexane.

**Synthesis of CdSe/CdS@Cd$_{1-x}$Zn$_x$S core/crown@shell CQWs:** Core/crown@shell heterostructure was synthesized by following a previously reported c-ALD recipe with some modifications.[2] 4 mL of of NMF was added to the dispersion of CdSe/CdS core/crown CQWs in 1 mL hexane. Then, 40 µL of 40-48% aqueous solution of ammonium sulfide was added as sulfur shell growth precursor and stirred for 2 min. Growth of the sulfur layer was terminated by the addition of 1 mL acetonitrile and excess toluene (until the mixture became blurry) and the mixture was then centrifuged. The precipitate was dispersed in NMF and the same cleaning procedure was repeated to remove unreacted precursor. Finally, the CQWs were dispersed in 4 mL of NMF. Cation precursor was prepared by mixing solutions of 0.4 M $Cd(NO3)2.4H_2O$ and 0.4 M $Zn(NO3)2.6H_2O$ in NMF by desired volume fraction to obtain a gradient in concentration of cations through the layers. For the cation deposition step, 1 mL of cation precursor mixture was added to CQW dispersion in NMF and it was stirred for 45 min. The same cleaning step was repeated to get rid of excess precursors. To increase number of shells, the aforementioned steps were repeated until a desired number of shells was deposited. The fraction of the zinc to cadmium in cation precursors used is 50:50, 95:5, 95:5 and 99:1 for the 1st, 2nd, 3rd and 4th layer, respectively. Lastly, CQWs were precipitated and 5 mL of hexane and 100 mL of OLA were added on top of the precipitate and the mixture was stirred overnight. To rid excess ligands, the dispersion of CQWs was precipitated by addition of a tiny amount of ethanol and stored in hexane.

**Optical gain measurements**

A hollow, transparent fiber filled with CQW solution was fixed from one end such that it was parallel to the optical table. The sample was excited using femtosecond laser pulses generated by a Ti:Sapphire laser amplifier, with pulse width of ~110 fs and at a repetition rate of 1 KHz and a wavelength of 800 nm. These pulses were frequency-doubled using a nonlinear crystal. After the frequency-doubling, a short-pass filter was used to eliminate the remaining 800 nm beam. The excitation was focused onto the sample using a cylindrical lens so that the excitation stripe was collinear with the hollow fiber. The incident power was controlled using an adjustable attenuator, and the power was simultaneously measured using a 50:50 beam splitter on the path. The beam profile on the sample was determined using a CCD camera on the focal point and approximated as an ellipse with a long axis of ~4 mm and a short axis of 150 µm. For VSL measurement, a tunable slit was placed onto the path right before the sample to change the stripe length.

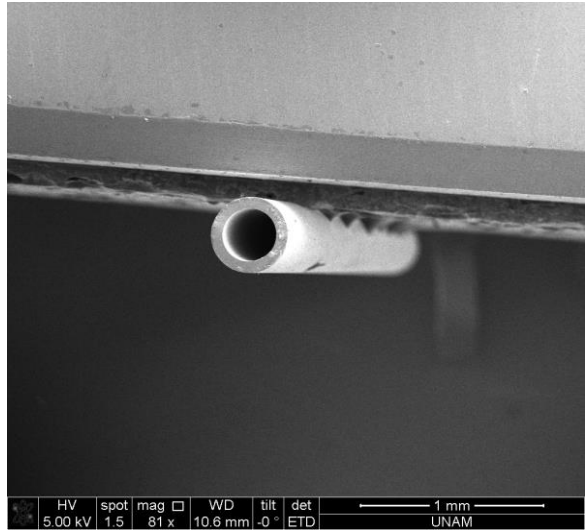

Figure S1. A SEM image of capillary tube used in this work.

**Numerical calculations for optical confinement factor**

Numerical computation was performed using MATLAB.[3,4] The waveguide was modeled by employing a step-index fiber approximation, assuming that the fiber with a core radius of 150 µm is surrounded by $SiO_2$ cladding extending to infinity. In addition, vertically and horizontally uniform illumination was assumed with a beam waist of 150 µm which was experimentally determined using a CCD camera. This model does not include all the bound modes as most of the bound modes are actually modes trapped in the $SiO_2$ cladding. These cladding modes have almost no overlap with the gain medium. Instead of calculating $\Gamma$ for all the bound modes, we found a maximum for the confinement factor and therefore a minimum for the gain cross section. A more detailed analysis requires calculating modal parameters and power distributions for more than hundred thousand modes which in our case is unnecessary. The effect of diffraction along the waveguide and nonuniform coupling to different modes were bundled as optical loss and not further analyzed. The optical constants for $SiO_2$ were taken from literature[5] whereas the core region consisting of CQW solution was modeled as a lossless/gainless dielectric material with an effective refractive index of 1.51. The refractive index was calculated using Bruggeman's effective medium theory (EMT):

$$\sum_i f_i \frac{\varepsilon_i + <\varepsilon>}{\varepsilon_i + 2<\varepsilon>} = 0, \quad \sum_i f_i = 1 \tag{S1}$$

where $\varepsilon_i$ is the complex dielectric constant of every material, $f_i$ is the filling factor of the corresponding material, and $<\varepsilon>$ is the effective dielectric function of CQW solution.[6] In our calculation we used the experimentally measured dielectric constants of all constituting materials and the average dimensions of CQWs were determined using transmission electron microscopy image analysis. The profile height parameter $\Delta$ was calculated as 0.12, whereas the fiber parameter $V$ of the resulting waveguides was calculated as 680 and 556 for 534 nm and 653 nm, respectively. $V$ values indicate that the guiding is extremely multi-moded, enabling the use of the approximations for modal parameter $U$.[7] Also, a low value for $\Delta$ shows that the fiber is in its weakly-guiding regime and its supported modes are TEM-like.[3] For the first couple of modes, the solutions for $U$ approximate the zeros of Bessel function of the first kind, $J_\nu(U)$. Modal power profiles for various TE, TM and HE modes were calculated and TE and TM modes were approximated as $HE_{2m}$ modes. HE modes prove the validity of weak-guiding approximation with TEM-like angle-independent power distribution. The confinement factors for the specified modes were calculated based on our definition given by the equation (3) in the main text.

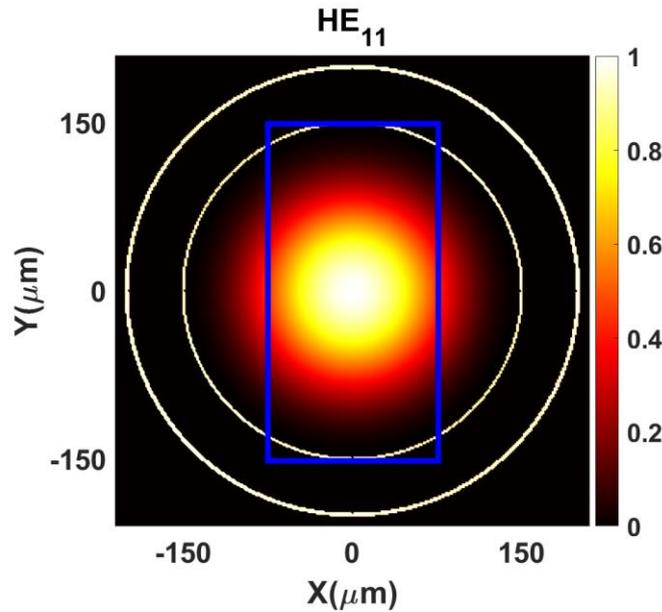

Figure S2. Spatial regions used for $\Gamma$ calculations. The blue rectangle shows the cross-section of the excitation. Large spatial overlap between HE$_{11}$ mode and excitation can be clearly seen. Illuminated width was measured as 150 µm using a CCD camera.